\documentstyle[12pt]{article}
\begin{document}

\title{\bf BOX ANOMALY AND $\eta' \to \pi ^+ \pi ^- \gamma$ DECAY}
\author{A.V. Kisselev and V.A. Petrov \\
\small 142284 Institute for High Energy Physics, Protvino, Russia}
\date{}
\maketitle

\begin{abstract}
In the framework of the "current scheme" of $\eta - \eta'$  mixing 
we calculate the amplitude  of the decay $\eta' \to  \pi ^+ \pi ^- 
\gamma $ in the soft limit  and  compare  our results with recently 
reported  values  of  the phenomenological additive terms to the 
$\rho $--meson background.
\end{abstract}

\section{Introduction}

In the experiment on the  study  of $\eta '  \to   \pi ^+  
\pi ^-  \gamma$  decay 
completed with a high accuracy by the LEPTON F Collaboration~\cite{Bityukov} 
it was found that  certain  discrepancy 
between the data and fits of the dipion spectrum based  solely  on 
the  $\rho $--meson  decay  could  be  avoided   if   the   $\rho 
$--meson contribution was coherently supplemented by some constant  
additive term.
A   subseqent   thorough   analysis   of   the   world   data   at 
CERN~\cite{Benayoun} confirmed such  a  solution.

Recently the new measurements of the dipion mass distribution in
$\eta '  \to   \pi ^+  \pi ^-  \gamma$ by the Crystal Barrel 
Collaboration~\cite{Abele} have also shown
the existence of a non--resonant contribution to this decay. 
The analysis of the decay $\eta '  \to   \pi ^+  \pi ^-  
\gamma$ from Ref.~\cite{Ivanov} has confirmed this conclusion.

It has been known for a long time (see e.g.~\cite{Chanowitz}) that 
the amplitude of the decay $\eta ' \to  \pi ^+ \pi ^- \gamma 
$ is  related  to  the 
chiral  anomalies  (AVV,  AAAV).  In  Ref.~\cite{Chanowitz*}  (and 
references therein)  this  decay  was  also  considered  as  an 
additional way  of  checking  the  quark  charge  assignment  (the 
standard  fractional  charges~\cite{Gell-Mann}   versus   integral 
ones~\cite{Bogolyubov}).

One should note that in comparison with, say, $\gamma \gamma $ decays, 
the use 
of the soft limit of the anomalous term in  the  lowest  order  is 
complicated by the $\rho $--meson dominance (responsible for the  bulk 
of the observed dipion mass spectrum).  In  Ref.~\cite{Chanowitz*} 
the  account  of  the  $\rho $--meson  was  made  by  multiplying  the 
low--energy   amplitude   by   the   Breit--Wigner   factor.    In 
Refs.~\cite{Benayoun}  it  was  assumed  that  this  is  the  term, 
additional to the $\rho $--meson term, which should be identified with 
anomalous contribution in the soft limit.  

The detailed analysis of the significance
of the $\rho$--resonance in $\eta'$ decay can be found in 
Refs.~\cite{Benayoun}.

In Ref.~\cite{Kisselev*} we evaluated the anomalous terms 
related to $\eta/\eta ' \to  \pi ^+ \pi ^- \gamma $   
decays in the soft limit for two $\eta - \eta'$ mixing 
schemes~\cite{Kisselev}.

In the present note we calculate the anomalous term in
$\eta ' \to  \pi ^+ \pi ^- \gamma$ in the "current mixing" scheme and 
compare our theoretical results with the experimental values reported in 
Refs.~\cite{Benayoun,Abele}.

\section{$\eta - \eta'$ mixing}

The amplitude we study contains anomalies that  come  through  the 
relationship of the Heisenberg field operators of $\eta $, $\eta '$ 
to the 
divergences of the $SU(3)$ octet and singlet axial  currents.  In 
Ref.~\cite{Kisselev}   the   following   expressions    for    the 
interpolating fields of $\eta $ and $\eta '$ were obtained
\begin{eqnarray}
\eta   &=&  \frac{1}{m_\eta ^2}  \left(  \frac{1}{F_8}  \cos   \theta    
D^8   - 
\frac{1}{F_0} \varepsilon \sin \theta  D^0 \right), \nonumber \\ 
\eta ' &=& \frac{1}{m_{\eta '}^2} \left( \frac{1}{\varepsilon F_8} 
\sin  \theta  
 D^8 + \frac{1}{F_0} \cos \theta  D^0 \right), \label{1}
\end{eqnarray} 
where $D_{8(0)}=\partial _\mu A^{\mu 5}_{8(0)}$ and $\langle \Omega |
A^{\mu 5}_{8(0)}|\eta (\eta ') \rangle = i 
p_\mu   f_{8(0)}$  ($F_{8(0)}=f_{8(0)} / \cos \theta $),  $A^
{\mu 5}_{8(0)}$ 
being octet (singlet) axial currents,  and  $\theta $  stands  for  
the mixing angle.
Equation~(\ref{1}) contains the factor  $\varepsilon $  which  is  
equal  to $m_\eta /m_{\eta '}$  in the so--called
"current mixing scheme" (see details in Ref.~\cite{Kisselev}).

As usual, when applying formulae like Eq.~(\ref{1}) we mean  the 
substraction of anomalous terms, so that in the soft limit Lorentz 
invariant formfactors of divergences disappear leaving us the net 
anomalous contribution (up to the sign).

\section{Basic formulae}

The amplitude of the decay $\eta' \to  \pi ^+ \pi ^- \gamma$ 
depends on the "decay constants" $f_{8(0)}$ and the mixing  
angle  which can be expressed in terms of the widths of $\gamma 
\gamma $ decays.
Equation~(\ref{1}) allows us to obtain the following relations
\begin{eqnarray}
R_\eta  &\equiv & \left[ \frac{3 \Gamma (\eta \to \gamma  
\gamma )}{\Gamma (\pi ^0 \to \gamma \gamma )}  \right]^{1/2} 
\left( \frac{m_\pi }{m_\eta } \right)^{3/2} = \frac{f_\pi }{F_8}  
\cos  \theta   - 
\sqrt{8} \, \frac{f_\pi }{F_0} 
\varepsilon \sin \theta , \nonumber \\
R_{\eta '} &\equiv & \left[ \frac{3 \Gamma (\eta ' \to  \gamma  
\gamma )}{\Gamma (\pi ^0 \to  \gamma  \gamma )} \right]^{1/2} 
\left(   \frac{m_\pi }{m_{\eta '}}  \right)^{3/2}   =    \frac
{f_\pi }{F_8} 
\frac{1}{\varepsilon} \sin \theta  + \sqrt{8} \, \frac{f_\pi }
{F_0}  \cos 
\theta. \label{2} 
\end{eqnarray} 

One can also relate the parameters $f_8$, $f_0$  and  $\theta $  to  the 
decays $J/\psi  \to  \eta \gamma $ and $J/\psi  \to  \eta '\gamma $.
Following~\cite{Novikov}, we suppose that the $J/\psi$ radiative decays are
dominated by $c \bar c$ annihilation into $gg \gamma$. The gluons $g$
are in the pseudoscalar state, and the ratio R of the widths of the decays
$J/\psi  \to  \eta \gamma $ and  $J/\psi  \to  \eta '\gamma$ is defined by
the ratio of the corresponding matrix elements of the gluon anomaly,
$\langle 0|G \tilde G |\eta' \rangle$, 
$\langle 0|G \tilde G |\eta \rangle$~\cite{Novikov}. The ratio R 
was estimated in Ref.~\cite{Kisselev}:
\begin{eqnarray}
R &\equiv&  \left[ \frac{\Gamma (J/\psi  \to  \eta ' \gamma)}{\Gamma (J/
\psi  \to \eta \gamma)} \right]^{1/2}  \left(
\frac{p_{\eta '}}{p_\eta }  \right)^{3/2}   \left( \frac{m_\eta }
{m_{\eta '}} 
\right)  ^2  = \left| \frac{\langle 0|G \tilde G |\eta'
\rangle}{\langle 0|G \tilde G |\eta \rangle} 
\right| \left( \frac{m_\eta }{m_{\eta '}} \right)  ^2 \nonumber \\
&=& \varepsilon  \frac{\sqrt{2}  \,  f_0  \cos   \theta    + 
\varepsilon f_8 \sin \theta }{- \sqrt{2} \, f_0 \sin \theta  + 
\varepsilon f_8 \cos \theta }, \label{3}
\end{eqnarray}
where $p_{\eta '}/p_\eta =(1-m^2_{\eta '}/m^2_{J/\psi })/(1-m^2_\eta /
m^2_{J/\psi })$.

One can thus express the mixing angle and  the  "decay 
constants" in terms of $R_{\eta} $, $R_{\eta '}$ and $R$. 
This was done and discussed in Ref.~\cite{Kisselev}. Experimental data  on 
the   decays   $\eta /\eta' \to \gamma \gamma $ and $J/\psi 
\to \eta /\eta' \gamma$~\cite{PDG} allow the 
production of the numerical results summarized in Table~1.
\vspace{0.5cm}

Table~1. Mixing angle and "decay constants" extracted from
$\eta / \eta' \to \gamma \gamma$ and $J/\psi \to \eta /\eta'\gamma$
decay.

\begin{center}
\begin{tabular}{c|c}
\hline \hline
$\theta ^\circ$ &  $- 19.62 \pm 2.33$ \\ \hline
$f_8/f_\pi $    &  $\hphantom{1}0.84 \pm 0.05$ \\ \hline
$f_0/f_\pi $    &  $\hphantom{1}0.88 \pm 0.07$ \\ 
\hline \hline
\end{tabular}
\end{center}
\vspace{0.5cm}

The  value  $\theta =(-19.7 \pm 2.2)^{\circ}$ is  in 
good agreement with the estimate $\theta  \approx - (19 \div 20)^{\circ}$ 
found in Refs.~\cite{Donoghue}. 

We also get
\begin{eqnarray}
\tan \theta  &=& [2 \varepsilon (R_{\eta} - 4 R_{\eta'} R)]^{-1}
\{ 5(\varepsilon^2 R_{\eta'} + R_{\eta} R) - \nonumber \\ 
&-& [9(R_{\eta} R - \varepsilon^2 R_{\eta'})^2 
+ 16(R_{\eta}^2  +  \varepsilon^2  R_{\eta'}^2) 
(\varepsilon^2 + R^2)]^{1/2} \}, \label{4}
\end{eqnarray} 
\begin{eqnarray}
\frac{f_\pi }{f_8} &=& R_\eta  + \varepsilon  R_{\eta'} \tan 
\theta , \nonumber \\
\frac{f_\pi }{f_0} &=& \frac{- R_\eta \tan \theta  + 
\varepsilon R_{\eta'}}
{\sqrt{8} \, \varepsilon}. \label{5}
\end{eqnarray} 
Note that $\theta $ depends only on $R$ and the ratio $R_\eta 
/R_{\eta'}$.

One should note that our formulas for the decay width ratios, underlying
the estimate of $\eta$--$\eta'$ mixing angle and the decay constants, 
differ from conventionally used formulas (see Refs.~\cite{Benayoun} and 
references therein). The latters can be reproduced in our approach by 
means of the formal substitution $\varepsilon=1$ in 
Eqs.~(\ref{1})-(\ref{5}).

\section{Renormalization scale dependence}

We now make some digression to discuss  the  question  of  the  RG 
properties of the "decay constants"  $f_0$  and  $f_8$.  Since  we 
consider QED in the  lowest  orders  no  question  arises  on  the 
renormalization scale dependence  of  both  anomalous  divergences, 
$D_8$ and $D_0$. We however cannot  neglect  QCD  renormalization. 
"Hard" nonconservation (anomaly)  of  the  singlet  axial  current 
leads  to  its  RG  scale  dependence   (corresponding   anomalous 
dimension  was  calculated  in  Ref.~\cite{Kodaira})  and,  as   a 
consequence, to the scale dependence of $D_0$ and $f_0$.

Due to multiplicative renormalization the matrix  element  of  the 
"decay" of $D_0$ in $\pi ^+ \pi ^- \gamma $ can be  cast  into  the  
following form 
\begin{eqnarray}
& & \langle \pi (p_+) \pi (p_-) \gamma (k)|D_0(\mu ^2)|\Omega 
\rangle \nonumber \\
& &  \hphantom{11111} = Z(\mu^2/q^2) \langle \pi (p_+)  \pi (p_-)
\gamma (k)|D_0(q^2)|\Omega \rangle, \label{6}
\end{eqnarray}
where $q=p_+ + p_- + k$.
The same factor $Z(\mu^2/q^2)$ renormalizes $f_0(q^2)$ to 
$f_0(m_{\eta'}^2)$, making the amplitude of the decay 
$\eta' \to \pi ^+ \pi ^- \gamma$ RG invariant (because $\eta'$ 
depends on the ratioo $D_0/f_0$ as one can see from Eq.~(\ref{1})).

The matrix element $\langle \pi \pi|D_0(q^2)|\Omega \rangle$ 
(with the pole at $m^2_{\eta'}$ subtracted)
is calculated in the soft limit $q^2 \to 0$, while $f_0(q^2)$ is taken 
at $q^2=m^2_{\eta'}$. It is usually assumed that one can put 
$\langle \pi \pi|D_0(0)|\Omega \rangle \simeq
\langle \pi \pi|D_0(q_0^2)|\Omega \rangle$, where $q_0 \simeq 1 
\mbox{ GeV}$. 
Note that due to the slow evolution of $Z(\mu^2/q^2)$ (the anomalous 
dimension starts from the second  order  in  $\alpha _s$~\cite{Kodaira})
the difference between $D_0(q_0^2)$ and $D_0(m_{\eta'}^2)$ is negligible.
So, one get $\langle \pi \pi|D_0(0)|\Omega \rangle / 
f_0(m_{\eta'}^2) \simeq \langle \pi \pi|D_0(m_{\eta'}^2)|\Omega \rangle / 
f_0(m_{\eta'}^2)$.

The RG properties of other matrix elements of $\eta'$ (the amplitude of
$\eta' \to \gamma \gamma$ decay, in particular) was analysed in 
Ref.~\cite{Kisselev**}.

\section{$\eta' \to \pi ^+ \pi ^- \gamma$ decay in the soft limit}

The amplitude of this decay has the  following  general  Lorentz 
structure
\begin{equation}
M(\eta' \to \pi (p_+)\pi (p_-)\gamma (k)) = E_P(p_+k, p_-k) \varepsilon _
{\mu \nu \rho \sigma } \varepsilon ^\mu k^\nu p^\rho _+  p^\sigma _-, 
\label{7}
\end{equation}
where  $\varepsilon ^\mu $ is the photon polarization vector.

Using Eq. (\ref{1}), and with all reservations made, we obtain the 
expression in the soft limit (both "box" and "triangle" anomalies 
contribute~\cite{Chanowitz}, ~\cite{Chanowitz*}):
\begin{equation}
E_{\eta '}(0) = - \frac{e}{4 \pi ^2 \sqrt{3} f_\pi ^2} \left(\frac{\sin 
\theta }{\varepsilon F_8} + \sqrt{2} \, \frac{\cos  \theta }{F_0}  
\right), \label{8}
\end{equation} 
where $e^2=4\pi \alpha_{em}$. 

With the parameters from Table~1 we find
\begin{equation}
E_{\eta '}(0) = -4.17 \pm 0.57. \label{9}
\end{equation}
In obtaining this estimate, we accounted for all the data on 
$\eta \to \gamma \gamma$~\cite{PDG} (with Primakoff--production 
measurements included). If we use only two--photon data, we get 
somewhat lower value of $E_{\eta'}$:
\begin{equation}
E_{\eta '}(0) = -4.01 \pm 0.38. \label{10}
\end{equation}
Equations~(\ref{9}), (\ref{10}) is the main result of our paper.

In Refs.~\cite{Benayoun,Abele} the phenomenological value $E_{\eta '}$ 
for the additive contribution (which according to  these  authors 
should be identified with $E_{\eta '}(0)$)  to  the  $\rho$ 
background depends on the choice  of  the  latter,  for  which  the 
freedom was reduced to two  variants:  model~1  (M1)  and  model~2 
(M2). More details  can  be  found  in  Refs.~\cite{Benayoun}.  
The corresponding results for $E_{\eta '}$ are  reproduced  
in Table~2.
\vspace{0.5cm}

Table~2. The experimental values of the additive  extra  terms  to 
the   resonance   background   for    two    variants    of    the 
latter.

\begin{center}
\begin{tabular}{c|c|c|c}
\hline \hline
 & \multicolumn{1}{c|}{\small Model M1} & 
   \multicolumn{1}{c|}{\small Model M2} & 
   \multicolumn{1}{c}{\small Ref.} \\ \hline
$E_{\eta'}$ & $- 5.06^{+ 0.53}_{- 0.54}$ & $- 2.17^{+ 0.49}_{- 0.46}$ 
& \cite{Benayoun} \\ \hline
$E_{\eta'}$ & $- 4.46 \pm 0.51$          & $- 1.78 \pm 0.53$ 
& \cite{Abele} \\ \hline \hline
\end{tabular}
\end{center}
\vspace{0.5cm}

Thus we have calculated the non--resonant contribution in 
$\eta' \to \pi ^+ \pi ^- \gamma$ decay, $E_{\eta'}$. 
The values obtained say definitely in favour of the model M1.
It is somewhat surprizing that our theoretical predictions obtained
in the soft limit, which is rather far from the physical one,
appears to be very close to the experimental values
of $E_{\eta'}$.


\end{document}